\documentclass[twocolumn,prc,aps,floatfix]{revtex4}
\usepackage[dvips]{graphicx}
\begin{document}
\title{Enhanced effect of temporal variation of the fine
 structure constant in diatomic molecules}
\author{V.V. Flambaum}
\affiliation{
 School of Physics, The University of New South Wales, Sydney NSW
2052, Australia
}
\date{\today}
\begin{abstract}
We show that the relative effect of variation of the fine structure constant
in microwave transitions between very close and narrow
rotational-hyperfine levels
may be enhanced 2-3 orders of magnitude in diatomic molecules
with unpaired electrons
like LaS, LaO, LuS, LuO, YbF and similar molecular ions.
 The enhancement is  result
of cancellation between the hyperfine and rotational intervals. 
\end{abstract}
\maketitle

 The possible variation of the fundamental constants
of nature is currently a very popular research topic.
There are  hints for the variation of the
fine structure constant, $\alpha=e^2/\hbar c$,  
strength constant of the strong interaction and masses
in Big Bang nucleosynthesis, quasar absorption spectra and Oklo
natural nuclear reactor data
(see e.g.\cite{var}) .
However, a majority
of publications report only limits on  possible variations
 (see e.g. reviews \cite{Uzan,karshenboim}).
 A very sensitive method to study the
 variation in a laboratory
 consists of the comparison of different optical and microwave atomic clocks
(see  recent measurements in
 \cite{prestage,Marion 2003,Bize 2005,Peik 2004,Bize 2003,
Fischer 2004,Peik 2005}).
An  enhancement of the relative effect of $\alpha$ variation can be obtained
in transition between the almost degenerate levels in Dy atom \cite{dzuba1999}.
These levels move in opposite directions if  $\alpha$ varies.
 An experiment is currently underway to place limits on
$\alpha$ variation using this transition \cite{budker}.
Unfortunately, one of the levels has  quite a large linewidth
and this limits the accuracy.

  In this paper I would like to note that very close narrow
levels exist in diatomic molecules. In the work \cite{khriplovich}
we suggested the use of such levels to enhance the effects of parity
 violation.
 Below I will show that there is
also an enhancement of  $\alpha$ variation in transitions between
such levels. 

  Consider, for example, the molecule $^{139}$La$^{32}$S (or LaO, LuO, LuS, YbF, etc, or similar molecular ions with  unpaired valence
 electrons).
The electron ground  state of this molecule is $^2\Sigma$.
The rotational-hyperfine levels are characterized by the
total angular momentum ${\bf J=L+F}$, rotational angular momentum $L$,
${\bf F=I+S}$, nuclear spin $I$ and electron spin $S$.
 The interval between the rotational-hyperfine levels
$|L=1,F=4,J>$ and $|L=2,F=3,J'>$ is close to zero within the experimental
accuracy (about 0.01 cm$^{-1}$, see all data in \cite{huber}).
 This is a result of the
cancellation between the rotational interval $B[L_1(L_1+1)- L_2(L_2+1)]$
($B$=0.11693 cm$^{-1}$) and the hyperfine interval $4b$ ($4b$=0.47 cm$^{-1}$);
the transition frequency is $\omega=4b-4B=0.002 \pm 0.01$ cm$^{-1}$. 

     The hyperfine structure constant can be presented in the following
form
\begin{equation}\label{A}
b=const \times [\frac{m_e e^4}{\hbar ^2}] [ \alpha ^2 F_{rel}( Z \alpha)]
[\mu \frac{m_e}{m_p}]
\end{equation}
The factor in the first bracket is the atomic unit of energy. The second 
``electromagnetic'' bracket determines the dependence on $\alpha$.
An approximate expression for the relativistic correction factor (Casimir
factor) for s-wave electron is the following
\begin{equation}\label{F}
F_{rel}= \frac{3}{\gamma (4 \gamma^2 -1)}
\end{equation}
where $\gamma=\sqrt{1-(Z \alpha)^2}$, Z is the nuclear charge.
Variation of $\alpha$ leads to the following variation of $F_{rel}$
 \cite{prestage}:
\begin{equation}\label{dF}
\frac{\delta F_{rel}}{F_{rel}}=K \frac{\delta \alpha}{\alpha}
\end{equation}
\begin{equation}\label{K}
K=\frac{(Z \alpha)^2 (12 \gamma^2 -1)}{\gamma^2 (4 \gamma^2 -1)}
\end{equation}
More accurate numerical many-body calculations \cite{dzuba1999}
 of the dependence of the hyperfine structure on $\alpha$ have shown
 that the coefficient $K$ is slightly larger than that given by this
formula. For Cs ($Z$=55) $K$= 0.83 (instead of 0.74),
  for Hg$^+$ $K$=2.28 (instead of 2.18).
 The last bracket in eq. (\ref{A})  contains the dimensionless
 nuclear magnetic moment $\mu$ in nuclear magnetons
 ( the nuclear magnetic moment $M=\mu\frac{e\hbar}{2 m_p c}$),
 electron mass $m_e$ and proton mass $m_p$. The dependence of $\mu$
on  the dimensionless strong interaction  parameter $X=m_q/\Lambda_{QCD}$
has been calculated in Ref. \cite{thomas,tedesco}. Here  $m_q$ is the quark
 mass and  $\Lambda_{QCD}$ is the strong interaction scale.

     The rotational constant can be presented in the following
form
\begin{equation}\label{B}
B=const \times [\frac{m_e e^4}{\hbar ^2}][\frac{m_e}{m_p}]
\end{equation}
Using eqs. (\ref{A},\ref{B}) we obtain  the frequency
\begin{equation}\label{omega}
\omega=const \times [\frac{m_e e^4}{\hbar ^2}][\frac{m_e}{m_p}]
[ \alpha ^2 F_{rel}( Z \alpha)\mu -const]
\end{equation}
We are interested in the case when two terms in the last bracket
 nearly cancel each other ($\omega=4b-4B \ll 4b$ ).
In this case $\omega$ is very sensitive to the variation of $\alpha$
and $\mu$.
The relative variation of the transition frequency is then approximately
 equal to 
\begin{equation}\label{delta}
 \frac{\delta \omega}{\omega} \approx \frac{4b}{\omega}
[(2+K) \frac{\delta \alpha}{\alpha}+\beta  \frac{\delta X}{X}]
\end{equation}
For La (Z=57) $K=0.9$ and $\beta \sim 0.01$ \cite{tedesco}.
 For a numerical estimate
we can assume the central value for the frequency,
 $\omega=4b-4B=0.002$ cm$^{-1}$. Then we have
\begin{equation}\label{deltan}
 \frac{\delta \omega}{\omega} \sim
600 \frac{\delta \alpha}{\alpha}
\end{equation} 
The accurate value of the coefficient
before $\delta \alpha /\alpha$ can be obtained after the frequency $\omega$
is measured.
Thus in molecular experiments one can have a two to three orders
of magnitude enhancement of the relative effect of  $\alpha$ variation.
Present laboratory sensitivity to the variation of $\alpha$ has reached
$2 \cdot 10^{-15}$ per year (see  review \cite{karshenboim}
and measurements \cite{Marion 2003,Bize 2005,Peik 2004,Bize 2003,
Fischer 2004,Peik 2005}). Thus, the required frequency stability
should be about 
\begin{equation}\label{delta1}
 \frac{\delta \omega}{\omega} \sim 10^{-12}
\end{equation}
For comparison, the microwave clocks have stability about  $10^{-16}$
(see e.g. \cite{Bize 2005}).
 It is also instructive to present an absolute shift of the frequency:
\begin{equation}\label{deltaomega}
\delta \omega \approx (2+K)4 b \frac{\delta \alpha}{\alpha}
\approx 4 \cdot 10^{10}  \frac{\delta \alpha}{\alpha} Hz
\end{equation} 
 This corresponds to  $\delta \omega
\sim 10^{-4}$ Hz per year.
The natural width of rotational levels in the ground state is small.
Therefore, the actual width and sensitivity to the variation of
$\alpha$ in molecules will depend on the particular experimental arrangement
which is beyond the scope of the present paper. Here  we only would like
to note that it was suggested in Ref. \cite{Cornell} to trap
molecular ions for the measurements of the electron electric dipole
moment. A similar trapping scheme may possibly be used to study
 the  variation of $\alpha$ in molecular ions.

This work is  supported by the Australian Research
Council.

\end{document}